\documentstyle[preprint,aps]{revtex}

\begin{document}

\begin{titlepage}
  
  \begin{center}
    \today

    \vspace {0.5cm}
    
    {\large \bf transport through \\ a constriction \\ in a FQH annulus}

    \vspace {0.5cm}

    Stefan Kettemann
    
    {\small \em 
Max- Planck Institute f.  Physik 
Komplexer Systeme,  Dresden, Germany  }

    \vspace {1cm}

  \end{center}

 The flux periodicity of thermodynamic properties of
an annulus in the fractional quantum hall state 
 with a constriction
  is considered.  It is found that $\phi_0$ - periodicity
 is obtained due to transfer of fractionally charged particles 
 or composite fermions between the edges of the annulus, respectively. 
 The result for the finite 
magnitude of the persistent current across a very strong
constriction is presented, as obtained with an extension of Wen's edge
 state theory.  

  \vspace {1cm}
  
  \noindent Keywords:  Fractional 
   quantum Hall effect, Persistent Current, Tunneling
  
  \vspace {1cm}

  {\small
    
  \noindent Corresponding Author: 
  
  Stefan Kettemann,

  \vspace {-0.3cm}

  Max- Planck Institute for Physics of Complex Systems,

  \vspace {-0.3cm}

  Noethnitzerstr. 38,

  \vspace {-0.3cm}

  01187 Dresden,

  \vspace {-0.3cm}

  Germany

  \vspace {-0.3cm}

  Fax: 49-351-871-1199

  \vspace {-0.3cm}

  e-mail: ketteman@mpipks.mpi-dresden.mpg.de }

\end{titlepage}

\newpage
\twocolumn

\section{Introduction}
 The Fractional Qunatum Hall
 effect, discovered by Stoermer, Tsui and Gossard, 
\onlinecite{fqh}  was explained by Laughlin in a 
  theory based on a  trial wave function\onlinecite{laughlin}.
 Since then, the 
question was debated if  quasiparticle 
excitations with fractional charge and statistics can be observed
 from such a state.

Thouless and Gefen considered 
 an  annulus  
 in a strong magnetic field
 which condenses the electrons to a fractional quantum hall state.
 They predicted   that 
thermodynamic properties of such a system have flux periodicity $\phi_0$
due to the existence of  families of 
states of the FQH- annulus. These  were connected by a physical mechanism,
the finite back scattering amplitude
of fractionally charged quasiparticles\onlinecite{5}.
 
Recently, this was verified 
by deriving the persistent current
in  a FQH- annulus with a weak constriction\onlinecite{me}.
 There, an extension of Wen's edge state theory\onlinecite{wen}
 was used, 
 which takes  into account the dynamics of zero- modes\onlinecite{haldane}.

 There exists the alternative theory of 
 the fractional quantum hall effect based on 
composite fermions
 as proposed by Jain \cite{jain}.
 
   A composite Fermion of a FQH state of filling factor $\nu = 1/(2 m+1)$
 has charge (-e), and 2m vortices attached to it, corresponding
 to 2 m flux quanta $\phi_0 = 2 \pi c/e$ aligned oppositely to the
 external magnetic field $B$.

\section{ The FQH annulus in the composite Fermion Picture}
   Let us consider the  annulus of a FQH liquid
 within the theory of composite fermions.
 When a 
 composite fermion is moved around a circle in the annulus, 
 which is pierced by an additional flux $\phi$,
it aquires
 a phase
\begin{equation} 
\chi =\frac{2 \pi}{\phi_0} \oint d r A = \frac{S B}{\phi_0}  + \frac{\phi}{\phi_0}
- 4 m \pi x,
\end{equation}
 where $S$ is the area enclosed by the circle\cite{gold}.
  x is the number of composite fermions within that circle.
 Since there are no composite fermions in the inner area $S_i$ of the
 annulus, we have $ x = \nu (S-S_i) B /\phi_0$, giving
\begin{equation}
\chi = 2 \pi \frac{ S \tilde{B}}{\phi_0} + 2 \pi \frac{\bar{\phi}}{\phi_0},
\end{equation}
 so that the composite fermion sees a reduced magnetic field 
\begin{equation}
 \tilde{B} = ( 1 - 2 m \nu ) B 
\end{equation}
 and an enhanced flux, piercing the annulus,
\begin{equation}
\bar{\phi} = \phi + 2 m \nu S_i B.
\end{equation} 
 We obtain the 
 guiding centers of the 
Eigenstates of the CF's by demanding quantization of the
 phase $\chi = 2 \pi l$, l an integer\cite{halperin}. 

  An adiabatic increase of the flux $\phi$ reduces the positions of the
 CF's. Thus,
  the FQH liquid moves up the inner edge.
  The area $S_i$ where there are
 no CF's becomes smaller, so that
 there are $\delta x_i$ CF's added to the inner edge with,
\begin{equation} 
\delta x_i = \nu \frac{ \delta \phi}{\phi_0}.
\end{equation}   
 An equal amount of CF's is removed from the outer edge.
  After a change of the flux by one flux quantum $\phi_0$,
 a fraction $\nu$ of a CF is added to the inner edge \cite{chk}.
   This is in agreement with the fact that   the Laughlin
 wave function yields the addition of a charge $- \nu e$ to the inner edge,
 when one flux quantum is added adiabatically into the annulus.
   If only electrons could tunnel between the edges of the 
 FQH annulus, its periodicity would be enhanced to $1/\nu \phi_0$.
  
 Is the  $\phi_0$- periodicity as required by the Byers- and Young
theorem \onlinecite{byers} restored
 by the transport of composite fermions between the edges?
  
 The removal of a CF from the inner edge adds 2m flux
 quanta to the flux piercing the annulus, since the two vortices attached
 to the removed electron did cancel 2 flux quanta of the magnetic field B.
  Thus, the total number of composite fermions added to the inner edge,
 when the flux piercing the annulus is changed by $\delta \phi$, 
 and one composite fermion tunnels to the outer edge, is:
\begin{equation}
\delta x_i = \nu \frac{\delta \phi + 2 m \phi_0 }{\phi_0} -1.
\end{equation} 
  Thus, after a flux change of 1 flux quantum,
 the tunneling of the composite fermion to the outer edge brought the system
 to its state at $\phi =0$, restoring  the $\phi_0$- periodicity.

 Therefore,  
in order that 
the flux periodicity of the FQH- annulus
 can be understood
  from the composite fermion perspective, 
 there has to be  tunneling of composite
 fermions between the edges of the annulus.

  It is now natural to ask if  the tunneling amplitude of
 composite fermions is renormalized,
 how this is compared with the renormalization of fractionally 
 charged particles\onlinecite{3}, and if it yields  the same
 temperature and size dependence of 
 the persistent current in the presence of a constriction as was obtained 
in Ref. \onlinecite{me}, using the edge state theory of fractionally charged
particles.
  To this end, one has to formulate the edge state theory of composite 
 fermions which is complicated by the dynamics of the vortices attached 
 to the fermions. Then, one can use this theory 
 to find a bosonized form of composite edge fermions.
 This is the subject of ongoing research.

  Here, rather we want to conclude by presenting results on 
 the persistent current through a very strong constriction using the 
edge state theory as presented in Ref. \onlinecite{me}.

\section{ Persistent current through a very strong constriction}

 We consider again an annulus in the FQH state at odd inverse filling factor.
   There is a strong constriction at one point in the annulus, producing a
   weak link through which particles can tunnel, only.
 The potential barrier 
 is taken to be  much larger than the energy gap in the bulk of the fractional
 quantum hall liquid,
$V_0 >> \Delta_{bulk} $.
 Then,  the many- body wave function of the FQH state in the annulus  
decays  exponentially.
 Thus,  the strong correlations 
 in the FQH state can not extend through it and 
only electrons can tunnel across the barrier.\\  

 We obtain the persistent current 
in second order perturbation theory in the most relevant electron tunneling 
amplitude $w_{1/\nu} $. We use the edge state theory of a single chiral edge
 connected at the constriction.
\begin{eqnarray}
&& I(\varphi) = - \frac{e}{\pi} \sum_{ang. mom. l} \partial_{\varphi} 
w_{1/\nu eff}( l, \varphi) 
\nonumber \\
&& \cos [ 2 \pi ( l + \varphi ) ], 
\end{eqnarray}
 where $\varphi = \phi/\phi_0$, $\phi$ being the flux 
 penetrating the annulus in addition to the constant background
 magnetic field B. . 
For temperatures below the level spacing,
 $T \ll v/L$,
 the renormalized tunneling amplitude of electrons $w_{eff}$ is given by
\begin{eqnarray}
&& w_{ 1/\nu} ( l , \varphi )_{eff} =
\nonumber \\ &&
 w_{ 1/\nu }(l, \varphi ) ( \pi v /(2 L
\Lambda ) )^{1/\nu}
\nonumber \\ &&
 ( 2 ( 1 - \cos ( \pi ( x_R - x_L )/L ) ) )^{-1/(2 \nu)}. 
\end{eqnarray}

   Here, L is the total length of the single chiral edge,
 and $x_R$ and  $x_L$ are the two positions at which the tunneling takes
 place. Their distance is taken to be of the order of $L/2$.
  $ w ( l, \varphi ) $ are the flux dependent coefficients of 
an expansion of  
the tunneling amplitude of electrons across the barrier in terms of the
 Eigen states of angular momentum $l$ of the annulus 
of noninteracting electrons without constriction.
 They satisfy the relation, $w_{1/\nu} ( l, \varphi + 1 ) = w_{1/\nu} ( l+ 1, 
\varphi ) $. 

 Thus, there   is 
in the presence of a very strong constriction 
a finite $ \phi_0$- periodic  persistent current
 of electrons 
across the weak link
 and its 
     amplitude decreases with increasing 
 circumference 
 L of the annulus only slowly, like a power law.

 When the temperature exceeds the level spacing,  $T \gg v/L$,
 the effective tunneling amplitude is 
\begin{eqnarray}
&&w_{ 1/\nu eff} ( l , \varphi ) = 
\nonumber \\ &&
w_{1/\nu} ( l, \varphi)  ( \pi T/\Lambda )^{1/\nu} 
\exp [ - 1/\nu T L/v 
\nonumber \\ &&
( 1 - \exp ( - \pi \mid x_R - x_L \mid /L )  ) 
\cos ( \pi v/( 2 T L ) )  ]
\nonumber \\ &&
 ( 1 - \exp ( - 2 \pi T/v  
 \mid x_R - x_L \mid  ) )^{- 1/\nu}
\end{eqnarray}  
 and the persistent current  has an  exponentially small amplitude. 
  
The derivation, as well as the result for a potential barrier smaller
 than the bulk energy gap, $V_0 < \Delta_{bulk} $ will be presented
 elsewhere.

\section{Conclusion} 

 The flux periodicity of a FQH annulus can be understood within the 
 picture of composite fermions.  The flux periodicty  $ \phi_0$ 
 is restored due to the existence of back scattering of composite fermions 
  between the edges. 

 There is a finite persistent current of electrons across a 
 very strong constriction, when the barrier height exceeds the bulk energy
 gap, at temperatures not exceeding the level spacing.

 The author would like to thank Yuval Gefen
 for many useful discussions and
 Konstantin Efetov and  Benham Farid for usefull comments on the
 tunneling through a barrier.
 
\onecolumn


\begin{thebibliography}{12}

\bibitem{fqh} H. L. St\" ormer, D. C. Tsui, A. C. Gossard, Surface Science
{\bf 113}, 32(1982),   
\bibitem{laughlin}  R.B. Laughlin, Phys. Rev. Lett. {\bf 50}, 1395(1983), 
\bibitem{5} D. J. Thouless, Phys. Rev. B {\bf 40}, 12034(1989), D. J. Thouless, Y. Gefen, Phys. Rev. Lett. {\bf 66} , 806(1991),
 Y. Gefen, D.J. Thouless   Phys. Rev. B {\bf 47}, 10423(1993
),
\bibitem{me} S. Kettemann,  Phys. Rev. B{\bf 55}, 2512-22(1997),
\bibitem{wen}  X. G. Wen, Phys. Rev. B {\bf 43}, 11025(1991), Int. J. Mod. Phys. B {\bf 6}, 1711(1992),
\bibitem{haldane} F. D. M. Haldane, J. Phys. C {\bf 14}, 2585(1981),
\bibitem{jain} J. K. Jain, Phys. Rev. Lett. {\bf 63}, 199(1989),
\bibitem{gold} A. S. Goldhaber, J. K. Jain, unpublished (1995),
\bibitem{halperin} B. I. Halperin, Phys. Rev. B {\bf 25}, 2185(1982).
\bibitem{chk} D. B. Chklovskii, B. I. Halperin, unpublished (1997),
\bibitem{byers} N. Byers and C. N. Young, Phys. Rev. Lett. {\bf 7},46(1961),
\bibitem{3} C. de C. Chamon and X. G. Wen, Phys. Rev. Lett. {\bf 70},
2605(1993), 
K. Moon, H. Yi, C. L. Kane, S. M. Girvin, and Matthew P. A. 
Fisher  
, Phys. Rev. Lett. {\bf 71}, 4381(1993),
\end{thebibliography}
\end{document}